%% file: main.tex
\documentclass[a4paper,11pt]{article}
\pdfoutput=1 

\usepackage{jinstpub} 

\usepackage{lineno}

\usepackage{booktabs, multirow} 
\usepackage{soul}
\usepackage[table]{xcolor} 
\usepackage{changepage,threeparttable} 

\title{\boldmath 
Embedded FPGA Developments in 130nm and 28nm CMOS for Machine Learning in Particle Detector Readout 
}

\author[1]{J. Gonski,}
\author[1]{A. Gupta,}
\author[1,2]{H. Jia,}
\author[1]{H. Kim,}
\author[1]{L. Rota,}
\author[1]{L. Ruckman,}
\author[1]{A. Dragone,}
\author[1]{and R. Herbst}
\affiliation[1]{
    SLAC National Accelerator Laboratory,
    2575 Sand Hill Road, M/S 96,
    Menlo Park, CA 94025, USA
}
\affiliation[2]{
    Stanford University, 
    450 Jane Stanford Way, 
    Stanford, CA 94305, USA 
}

\emailAdd{jgonski@slac.stanford.edu}

\abstract{

\noindent 

Embedded field programmable gate array (eFPGA) technology allows the implementation of reconfigurable logic within the design of an application-specific integrated circuit (ASIC). 
This approach offers the low power and efficiency of an ASIC along with the ease of FPGA configuration, particularly beneficial for the use case of machine learning in the data pipeline of next-generation collider experiments. 
An open-source framework called "FABulous" was used to design eFPGAs using 130 nm and 28 nm CMOS technology nodes, which were subsequently fabricated and verified through testing.
The capability of an eFPGA to act as a front-end readout chip was assessed using simulation of high energy particles passing through a silicon pixel sensor. 
A machine learning-based classifier, designed for reduction of sensor data at the source, was synthesized and configured onto the eFPGA. 
A successful proof-of-concept was demonstrated through reproduction of the expected algorithm result on the eFPGA with perfect accuracy. 
Further development of the eFPGA technology and its application to collider detector readout is discussed.
}

\keywords{
Reconfigurable Computing;
Open Source;
FPGA;
ASIC;
Machine Learning
}

\arxivnumber{2404.17701} 

\begin{document}
\maketitle
\flushbottom
\setlength{\parskip}{0pt}


\input{physics_motivation}

\input{cmos130nm}
\input{cmos28nm}
\input{ml_on_eFPGA}
\input{summary}

\acknowledgments

This work was supported by the U.S. Department of Energy under contract number DE-AC02-76SF00515.

\input{bibliography}
\end{document}

%% file: physics_motivation.tex
\section{Background}
\label{sec:Background}

\noindent 


Silicon microelectronics technology is a key part of the modern computational paradigm, offering high performance and low power options for data processing tasks.
A diversification of microchip designs offers the ability to customize computing solutions to a particular problem.
Common options such as application-specific integrated circuits (ASICs) or field-programmable gate arrays (FPGAs) offer trade-offs between performance, power consumption, and flexibility to suit a wide variety of applications. 
While an FPGA is fully customizable, it draws more power than an ASIC; however, the ASIC requires considerable expertise to design, and is fixed to a certain task once fabricated. 
Microelectronics remain a rapidly developing field, with advances promising to deliver even denser logic, faster processing, and novel designs.

In parallel, artificial intelligence and machine learning (AI/ML) have become essential tools in data science. 
In light of increasing dataset sizes and expanding computational capacity, ML-based methodology can execute common scientific tasks such as signal-noise classification, regression of key quantities, fast generation of simulation, or anomaly detection, with good efficiency and performance.  
Leveraging ML in these scenarios requires a high degree of reconfigurability in computing architecture, such that weights and biases of a model, as well as the model itself, can be updated throughout training and over time as the task evolves. 

In the sciences, many data processing tasks have a requirement to perform sophisticated AI inference within resource-constrained systems, such as with low power or latency, motivating the incorporation of ML into microelectronics.
This is enabled by ``fast ML", wherein high-level synthesis of software algorithms enables the implementation of ML models in hardware. 
In this work the \textit{embedded FPGA} (eFPGA) technology is studied as an option to dynamically reconfigure logic to enable ML within an ASIC. 
eFPGAs comprise a digital block of FPGA fabric that can be embedded into an ASIC design. 
In this way they can offer the best of both worlds between FPGA and ASIC processing, with lower power demand than an FPGA but with the ease of firmware-based reconfiguration. 
eFPGAs also mitigate issues of commercial FPGA obsolescence by providing a hardware structure that can be maintained without industrial support, making it easier for computational systems to be maintained over the long lifetime of a scientific experiment.

In high energy collider facilities like the Large Hadron Collider (LHC) at the CERN laboratory in Geneva, Switzerland, the ability to execute reconfigurable and fast logic is indispensable. 
The LHC collides bunches of protons at a frequency of 40 MHz, creating thousands of particles that pass through detectors surrounding the beam interaction point. 
These conditions create a requirement to handle a total data rate of hundreds of terabytes of data per second, with latencies as low as $\mathcal{O}$(ns). 
Furthermore, the high radiation doses and stringent spatial constraints associated to the collision environment motivate the development of custom ASICs and electronics to read out the detector channels at high speed with good fidelity. 

Fast ML is being explored in the collider context and across the sciences~\cite{fastMLscience}.
A common application for fast ML at colliders is with the trigger system, which comprises hardware- and software-based algorithms that decide which collision events should be written to disk based on real-time assessment of potentially interesting physics activity~\cite{atlas_trigger, cms_trigger}. 
ML in the trigger systems of LHC experiments is being studied to perform fast featurization~\cite{cms_phase2} or anomaly detection~\cite{40MHz} on incoming data.
Before the trigger, front-end on-detector readout electronics take low-level information collected by detector elements, perform simple operations such as amplification or digitization, and transmit this data off-detector for further processing. 
Information collected by the front-end readout electronics ultimately feeds the trigger algorithm and the eventual data acquisition, making this information an essential deliverable of the detector. 

Existing work to implement ML in the front-end focuses on dedicated ASIC designs for a particular architecture or model~\cite{MLfeASIC, neuromorphic}, which, while functional, is limiting in scope due to the rigidity of the hardware. 
eFPGAs can fill a crucial technology gap for high energy physics (HEP) experiments by providing an ASIC that can allow not only the reconfiguration of ML model weights, but the entire ML architecture itself onto the ASIC. 
As the transmission of data off the detector is often less efficient than data processing, the performance of sophisticated data compression and processing as close to the front-end as possible is a key task for modern collider experiments. 
Such capability will become crucial for the more challenging environments of future colliders~\cite{p52023}, where experiments will have to cope with data rates on the exascale with even higher radiation doses and limited space for material budget or cooling services.  
As a result, community driven exercises on detector R\&D for HEP list increased intelligence on-detector as a key priority for the success of future detector upgrades and designs~\cite{doe_brn,ushiggsfactory}. 

This work describes the design and fabrication of eFPGAs using the 130nm and 28nm CMOS technology nodes.
The open-source "FABulous" framework~\cite{woset_2021} is used to generate the eFPGA fabric.
An application of the 28nm eFPGA for at-source ML-based classification is performed, using data from simulated silicon pixel sensors in a high energy collider detector. 
The unique features of the eFPGA, namely its high degree of flexibility, relatively low power and footprint, and publicly accessible design framework make it an ideal tool for known data acquisition challenges in collider physics.

%% file: cmos130nm.tex
\section{Developments in 130nm CMOS}
\label{sec:cmos130nm}
\subsection{eFPGA Customization}
\label{sec:cmos130_eFPGA_customization}
\begin{figure}[tb]
\centering
\includegraphics[width=1.0\textwidth]{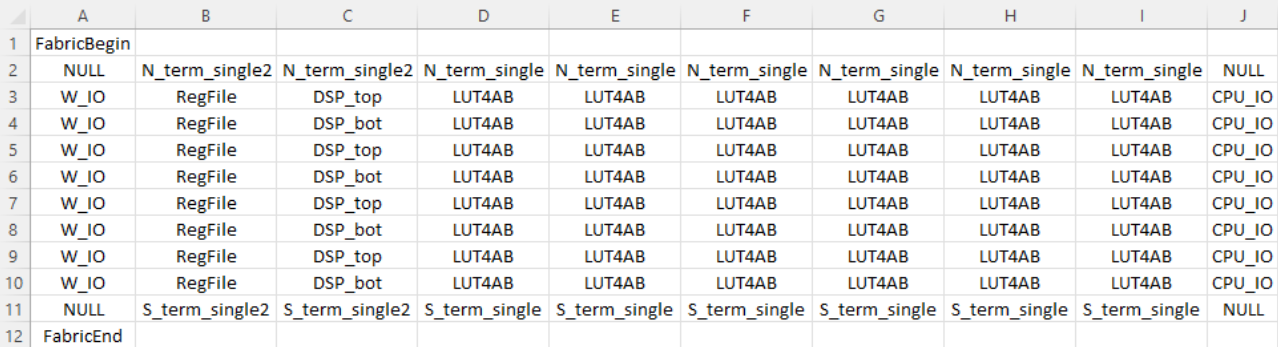}
\caption{\label{fig:cmos130_efpga_tile_config} Screenshot of the 130nm eFPGA tile configuration file.}
\end{figure}

The ability to customize the tiles of the eFPGA is provided by the FABulous framework~\cite{woset_2021}. 
Because this was our very first FABulous eFPGA to design, extreme conservatism was practiced, and the reference design provided by FABulous, known as ``fabric\_TSMC\_example", was utilized, which has been proven suitable for tapeout. 
A \textsc{.csv} file is used by FABulous framework to customize the eFPGA tile configuration, which is shown in Figure~\ref{fig:cmos130_efpga_tile_config} for the 130nm eFPGA. 
The eFPGA tiles utilized in this design included:
\begin{itemize}
   \item W\_IO: A 2-bit general-purpose input/output (GPIO) interface with tri-mode support (input, output, and high-impedance state)
   \item RegFile: A simple dual-port LUTRAM, consisting of 32 entries of 4 bits
   \item DSP\_top \& DSP\_bot pair: Digital signal processor (DSP) primitive, providing 8x8 multipliers with 20-bit accumulators
   \item LUT4AB: A logic cell comprising a Look-Up Table (LUT) with 4 inputs and a Flip-Flop (FF)
   \item CPU\_IO: An interface of 8-bits per tile from CPU to eFPGA, and 12-bits per tile from eFPGA to CPU
   \item NULL: An empty tile, used for terminating at eFPGA corners
   \item N\_term\_single2: Used for terminating the column on the northern side
   \item s\_term\_single2: Used for terminating the column on the southern side
\end{itemize}
The total resources for the 130nm eFPGA consisted of 384 logic cells, 128 registers, and 4 DSP slices.

\subsection{ASIC Digital Architecture}
\label{sec:cmos130_digital_arch}
\begin{figure}[tb]
\centering
\includegraphics[width=0.75\textwidth]{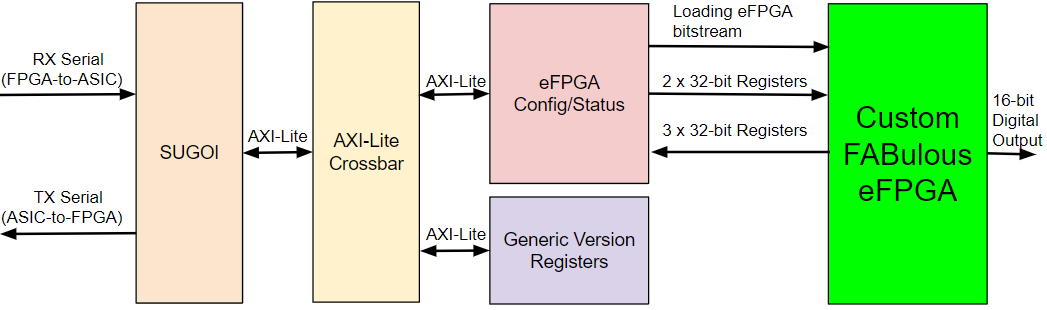}
\caption{\label{fig:cmos130_asic_block} Block diagram of the 130nm CMOS ASIC design.}
\end{figure}

A block diagram of the 130nm ASIC's digital architecture is shown in Figure~\ref{fig:cmos130_asic_block}. For register access from an external FPGA to this ASIC to be performed, a SLAC Ultimate Gateway Operational Interface (SUGOI) digital block is utilized. SUGOI is a simple packet-based control protocol that allows memory-mapped resources within ASICs to be read and modified using 8B10B serial encoding~\cite{sugoi}. Communication by the SUGOI module is carried out via an AXI-Lite~\cite{axil} crossbar to two separate AXI-Lite endpoints: an eFPGA configuration/status module and a generic version registers module. Diagnostic information about the ASIC, such as the git hash of the code when the ASIC was taped out and the revision number of the ASIC, is provided by the generic version registers module. The eFPGA status/configuration module is employed for loading the bitstream into the eFPGA itself, for reading from multiple 32-bit buses of the eFPGA, and for driving multiple 32-bit buses to the eFPGA. A 16-bit digital output bus via the W\_IO tiles for monitoring/debugging on a digital logic analyzer is featured by the eFPGA.

\subsection{Fabrication}
\label{sec:cmos130_fabrication}
\begin{figure}[tb]
\centering
\includegraphics[width=1.0\textwidth]{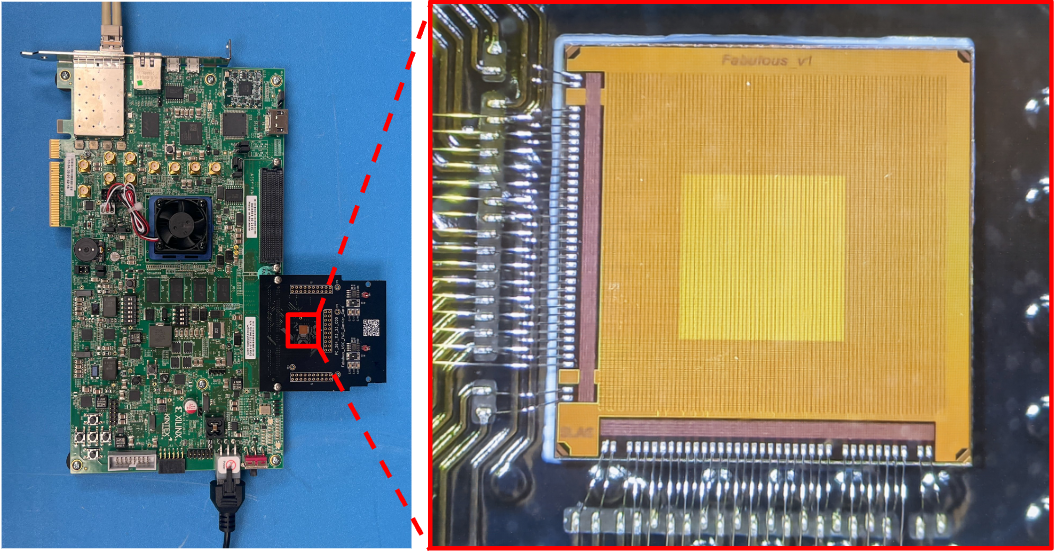}
\caption{\label{fig:cmos130_fmc_kcu105} (Left) Photograph of the KCU105 development board with the custom FMC ASIC carrier with ASIC wire bonded to it. (Right) Zoomed in photograph of the 130nm CMOS ASIC (5mm x 5mm).}
\end{figure}

The submission of this ASIC design to the Taiwan Semiconductor Manufacturing Company (TSMC) 130nm Multi Project Wafer (MPW) was completed in May 2022, and the design was received in August 2022. 
The design of the custom Printed Circuit Board (PCB) carrier adhered to the FPGA Mezzanine Card (FMC) form factor~\cite{fmc}. 
Firmware and software were developed to control and monitor the ASIC through an AMD KCU105 development board~\cite{kcu105}.
A photograph showing the 130nm CMOS ASIC wire-bonded to a custom PCB carrier can be seen in Figure~\ref{fig:cmos130_fmc_kcu105}, as well as a photograph displaying the ASIC on the FMC card alongside the KCU105. The dimensions of the custom PCB carrier are 6.90 cm x 7.65 cm.

\subsection{Testing Results}
\subsubsection{Simple Counter Test}
\label{sec:cmos130_simple_couner_test}

\begin{figure}[tb]
\centering
\includegraphics[width=1.0\textwidth]{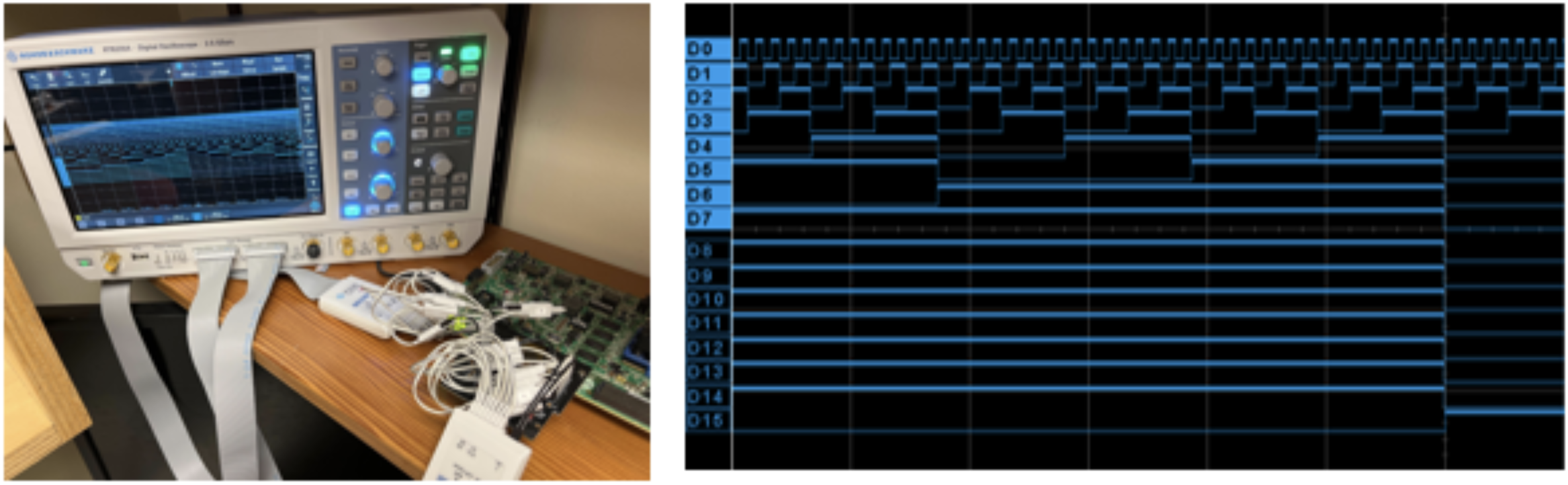}
\caption{\label{fig:cmos130_counter_test} Photograph of the 130nm ASIC connected to a logic analyzer (left), and logic analyzer display of the eFPGA loaded with a simple 16-bit counter bitstream (right).}
\end{figure}

The basic functionality of the eFPGA was evaluated through the compilation and loading of the bitstream containing simple 16-bit counter firmware into the eFPGA. 
A photograph showing the setup for this test can be seen in Figure \ref{fig:cmos130_counter_test}. 
The bus of the eFPGA counter was interfaced with the GPIO tiles, which subsequently interfaced with a 16-pin header. This header was then connected to a digital logic analyzer, facilitating the observation of the eFPGA 16-bit counter firmware's behavior. The behavior of the firmware was observed to be as anticipated, thus demonstrating the successful loading of the bitstream.

\subsubsection{ASIC Power Draw}
\label{sec:cmos130_power_draw}

\begin{figure}[tb]
\centering
\includegraphics[width=1.0\textwidth]{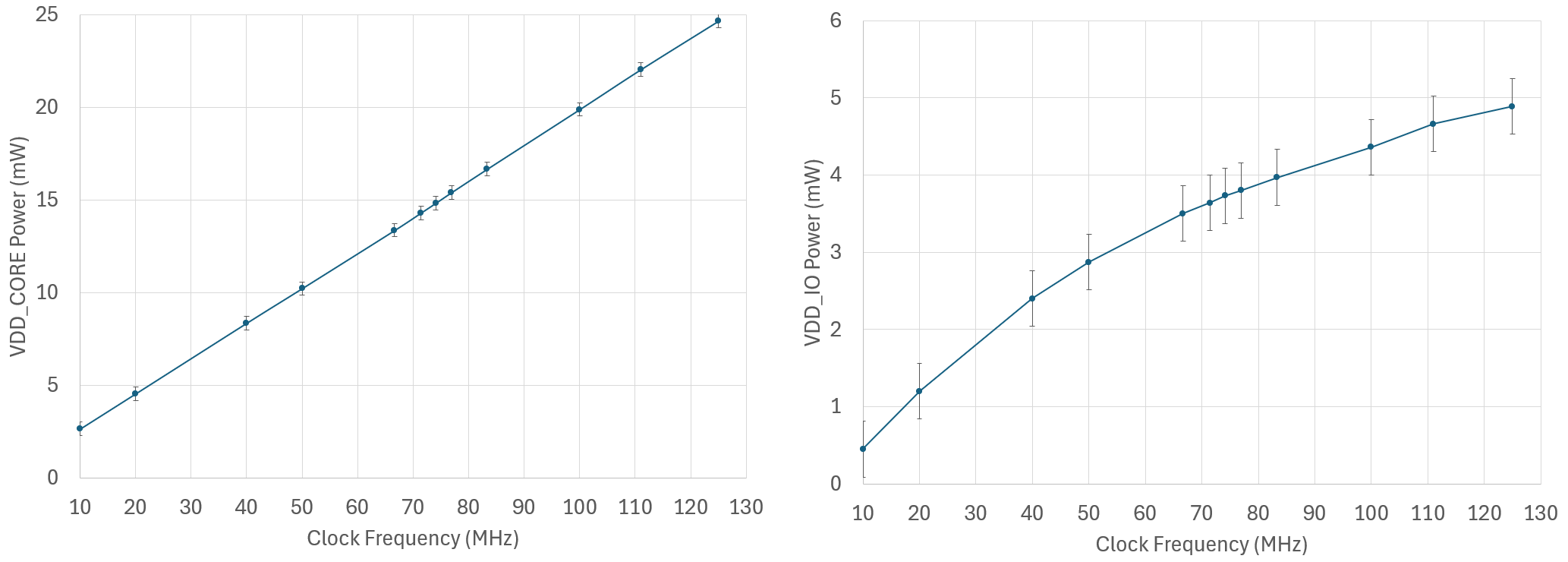}
\caption{\label{fig:cmos130_pwr_vs_freq} Plot of the 130nm ASIC core voltage power draw versus clock frequency (left), and plot of the 130nm ASIC I/O voltage power draw versus clock frequency (right).}
\end{figure}

To measure the power draw, the same firmware from the 16-bit counter test was used, and the clock frequency of the ASIC was varied. This clock generated by the KCU105's FPGA is sourced by the ASIC's SUGIO, AXI-Lite endpoints, and the eFPGA itself. 
At various clock frequencies ranging from 10 to 125 MHz, both the ASIC's core rail (+1.2VDC) and I/O rail (+1.2VDC) current draws were measured to calculate the total ASIC power drawing. Plots of these measurements are shown in Figure \ref{fig:cmos130_pwr_vs_freq}. The core voltage power consumption was observed to increase linearly with the clock frequency. This linear relationship is influenced by the frequency dependence of the switching events in the circuit. The non-linear power consumption for the I/O voltage is due to the load of the ASIC PCB carrier and KCU105's FPGA, which causes an incomplete settling of the digital signals at high frequencies (while still working correctly), thus reducing the current consumption.

When the clock frequency exceeded 74 MHz, a lock on the SUGOI link was still achieved by the ASIC, whereas the FPGA failed to achieve such a lock. The problem was traced back to the ASIC's output driver, which was determined to be excessively slow. The slew rate of this output driver was measured to be a rising edge of 38 ns and a falling edge of 32 ns. Conversely, the rising and falling edges for the FPGA were found to be on the order of a few nanoseconds. Beyond a clock frequency of 74 MHz, it was possible to only set the ASIC registers via the SUGOI interface, with the functionality to read them back being unavailable. Measurements were ceased after 125 MHz, as this was the timing constraint employed in the place and route software for the ASIC's digital logic.

%% file: cmos28nm.tex
\section{Motivation for Transitioning from 130nm to 28nm CMOS Technology}
The motivation for the transition from 130nm to 28nm CMOS technology was driven by the pursuit of increased logic density, enhanced radiation hardness~\cite{cmos28_rad_hard_A,cmos28_rad_hard_B,cmos28_rad_hard_C,cmos28_rad_hard_D}, and reduced power consumption. This shift was required to meet the evolving eFPGA-based data processing requirements. Through the utilization of a smaller CMOS technology node, significant improvements in the performance and adaptability of eFPGA architectures were made possible. The optimization of the physical characteristics of the eFPGA, to take advantage of the benefits offered by the 28nm technology, was the focus of the design choices. Specifically, the decreased feature size of the 28nm process allowed for the integration of more complex logic circuits at higher densities, resulting in a factor of 21 improvement in area efficiency and a factor of 2.8 reduction in core power consumption at 100 MHz.
Combined with the need to keep pace with foundry production capabilities, this development also ensures the eFPGA can keep pace with industry standards and be viable for large-scale production towards future collider detector needs.

\section{Developments in 28nm CMOS}
\label{sec:cmos28nm}
\subsection{eFPGA Customization}
\begin{figure}[tb]
\centering
\includegraphics[width=1.0\textwidth]{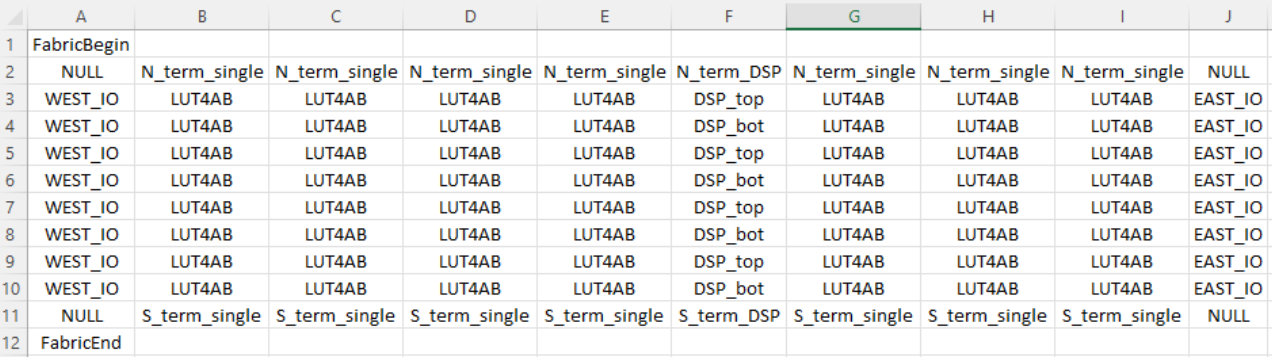}
\caption{\label{fig:cmos28_efpga_tile_config} Screenshot of the 28nm eFPGA tile configuration file.}
\end{figure}

The 28nm eFPGA design is similar to the 130nm eFPGA configuration described in Section~\ref{sec:cmos130_eFPGA_customization}. 
The primary distinction between the 28nm eFPGA and the 130nm eFPGA configuration lies in the removal of ``RegFile" tiles, which have been substituted by LUT4AB tiles. The integrated flip-flop included in the LUT4AB tiles is utilized for register functions. Additionally, the replacement of W\_IO tiles with WEST\_IO tiles, and CPU\_IO tiles with EAST\_IO tiles, is made.  
``User defined" tiles, namely WEST\_IO and EAST\_IO tiles, have been developed to enhance the interconnectivity between the eFPGA and the remaining digital logic within the ASIC. All the other tiles are defined in Section~\ref{sec:cmos130_eFPGA_customization}. 
The \textsc{.csv} file used by FABulous framework to customize the 28nm eFPGA tile configuration is shown in Figure~\ref{fig:cmos28_efpga_tile_config}. 
In total, the 28nm eFPGA comprises 448 logic cells and 4 DSP slices.

\subsection{ASIC Digital Architecture}
\begin{figure}[tb]
\centering
\includegraphics[width=0.75\textwidth]{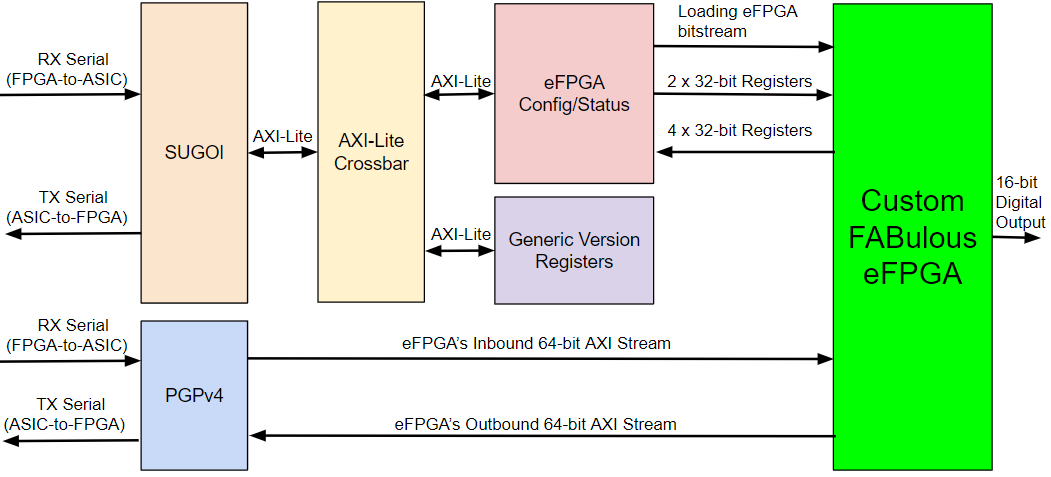}
\caption{\label{fig:cmos28_asic_block} Block diagram of the 28nm CMOS ASIC design.}
\end{figure}

A block diagram of the 28nm ASIC's digital architecture is shown in Figure~\ref{fig:cmos28_asic_block}. 
The digital architecture is very similar to the 130nm digital architecture described in Section~\ref{sec:cmos130_digital_arch}, with two major differences. 
The first is the number of 32-bit buses from the eFPGA to the eFPGA configuration/status module, which was increased from three to four.  
The second is the addition of AXI streams~\cite{axis} to/from the eFPGA. The AXI streams to/from the eFPGA are connected to a Pretty Good Protocol Version 4 (PGPv4) module. 
PGPv4 is a  low-latency, 64B66B-based serial protocol for high-speed data transfer over point-to-point link between FPGA/FPGA or FPGA/ASIC communication\cite{pgp4}.

\subsection{Fabrication}
\begin{figure}[tb]
\centering
\includegraphics[width=0.75\textwidth]{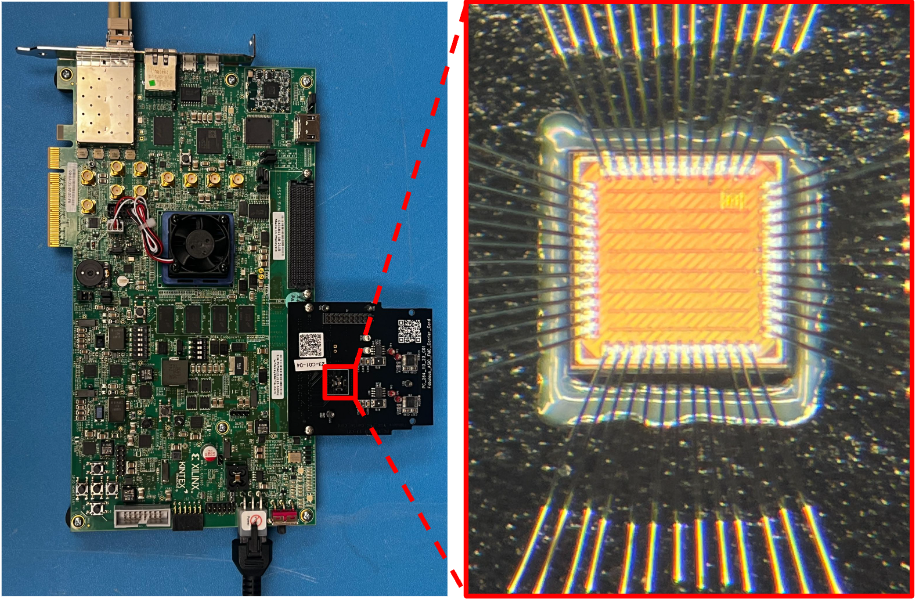}
\caption{\label{fig:cmos28_fmc_kcu105} (Left) Photograph of the KCU105 development board with the custom FMC ASIC carrier with ASIC wire bonded to it. (Right) Zoomed in photograph of the 28nm CMOS ASIC (1mm x 1mm).}
\end{figure}

The submission of this ASIC design to the TSMC 28nm MPW was completed in July 2023, and the design was received in January 2024. 
Although the custom PCB carrier for the 28nm ASIC differs from that of the 130nm ASIC, the majority of the firmware and software was adapted from the previous project with minor modifications to incorporate streaming PGPv4 support, utilizing the same KCU105 development board. 
A photograph showing the 28nm CMOS ASIC wire-bonded to a custom PCB carrier, and the ASIC on the FMC card alongside the KCU105, can be seen in Figure~\ref{fig:cmos28_fmc_kcu105}. The dimensions of the custom PCB carrier are 6.90 cm x 7.65 cm.

\subsection{Testing Results}
\subsubsection{Simple Counter Test}
\label{sec:cmos28_simple_couner_test}
\begin{figure}[tb]
\centering
\includegraphics[width=0.75\textwidth]{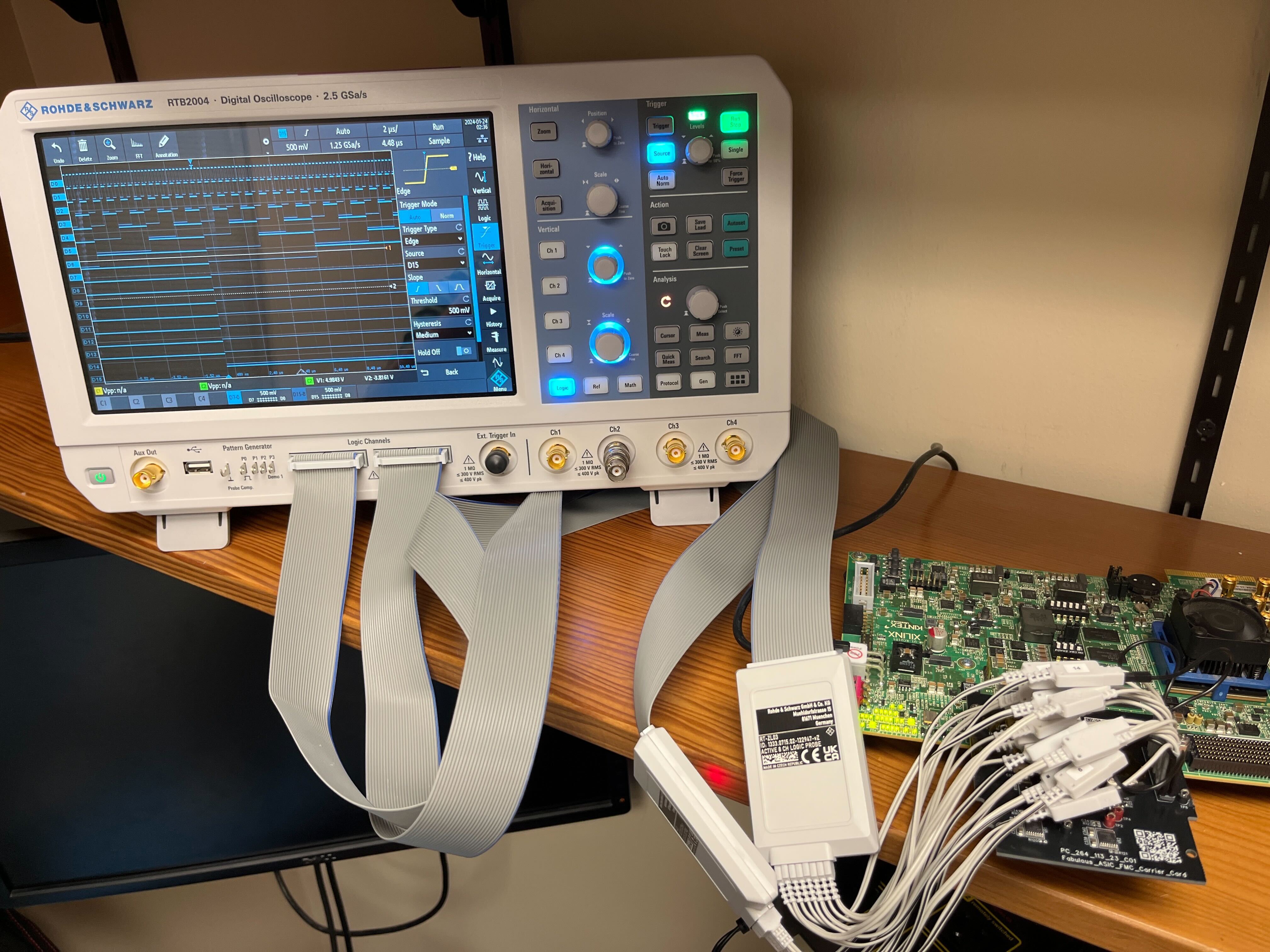}
\caption{\label{fig:cmos28_counter_test} Photograph of the 28nm ASIC connected to a logic analyzer with the eFPGA loaded with a simple 16-bit counter bitstream.}
\end{figure}

Similar to Section \ref{sec:cmos130_simple_couner_test}, the basic functionality of the eFPGA was evaluated through the compilation and loading of the bitstream containing very simple 16-bit counter firmware into the eFPGA. 
A photograph showing the setup for this test can be seen in Figure~\ref{fig:cmos28_counter_test}. 
In the absence of GPIO tiles, 16 bits from the WEST\_IO tiles were tapped off and driven off chip, which subsequently interfaced with a 16-pin header. This header was then connected to a digital logic analyzer, facilitating the observation of the eFPGA 16-bit counter firmware's behavior. 
As with the 130nm eFPGA, the behavior of the firmware was confirmed, indicating successful loading of the bitstream.

\subsubsection{ASIC Power Draw}
\label{sec:cmos28_power_draw}
\begin{figure}[tb]
\centering
\includegraphics[width=1.0\textwidth]{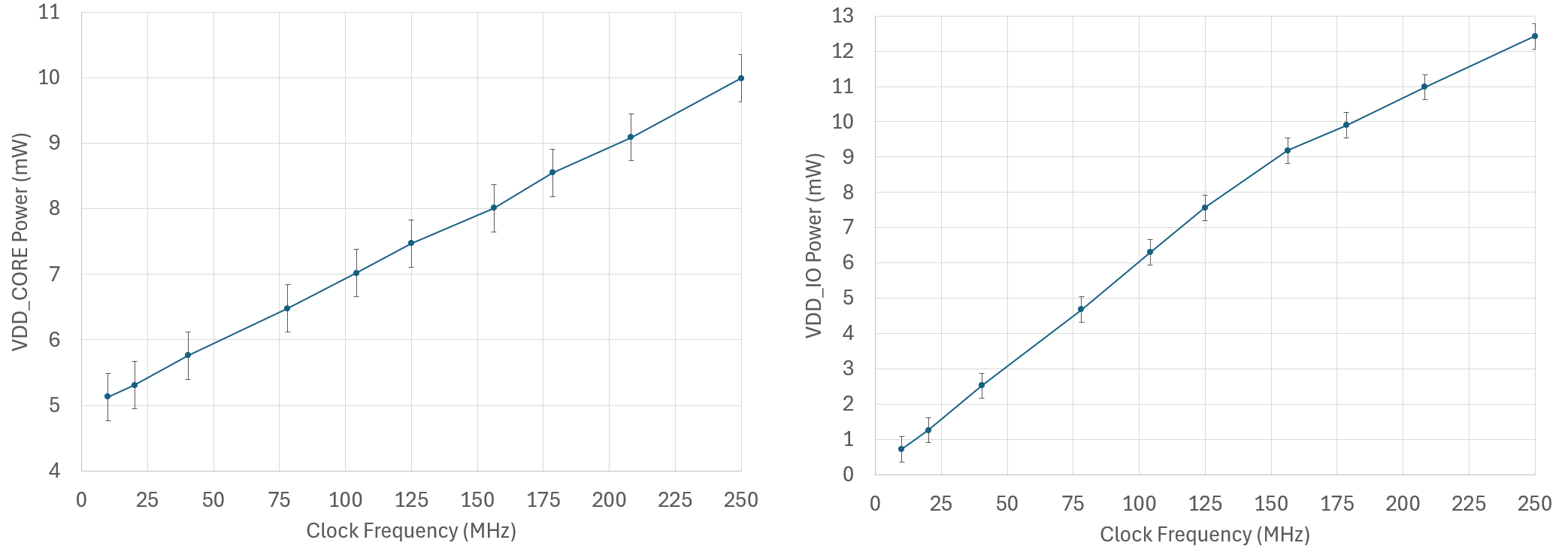}
\caption{\label{fig:cmos28_pwr_vs_freq} Plot of the 28nm ASIC core voltage power draw versus clock frequency (left), and plot of the 28nm ASIC I/O voltage power draw versus clock frequency (right).}
\end{figure}

Similar to the testing done at 130nm, the eFPGA 16-bit counter firmware was used for the 28nm ASIC, with variations in the clock frequency. The current draws from the ASIC's core rail (+0.9VDC) and I/O rail (+1.8VDC) were measured to calculate the total ASIC power consumption. For the 130nm ASIC, the same supply voltage was utilized for both the I/O circuitry and the core. In contrast, the 28nm ASIC employed distinct supply voltages for the I/O circuitry and the core. This choice was primarily influenced by the simplicity and availability of IP cores needed for interfacing the ASIC with the FPGA, but it did not significantly affect the key performance metrics.
The resulting measurements are shown in Figure \ref{fig:cmos28_pwr_vs_freq}, demonstrating that the 28nm ASIC's core voltage rail power consumption at a 125 MHz clock is approximately one third that of the 130nm ASIC design.

In contrast to the 130nm ASIC (refer to Section \ref{sec:cmos130_power_draw}), issues with the CMOS output driver slew rate of the 28nm ASIC were not observed and appeared to be comparable to that of the KCU105's FPGA. A stable SUGOI link lock was achieved on both the ASIC and FPGA sides from 10 MHz to 250 MHz. Although the timing constraints used in the place and route software for the ASIC's digital logic were set for 200 MHz (5ns clock period), no unusual behavior was noted. Measurements were discontinued beyond 250 MHz due to the FPGA's inability to achieve timing closure on a combinatorial chain in the PGPv4 protocol, which was related to calculating a 32-bit Cyclic Redundancy Check (CRC) value.

\subsubsection{AXI Stream Loopback in the eFPGA}
To test the eFPGA's AXI stream interface, the firmware was developed to enable loopback of the inbound stream into the outbound stream through a single register stage with back pressure handshaking being implemented. For the purpose of this testing, PseudoRandom Binary Sequence (PRBS) frames generated by the software were transmitted to the KCU105's FPGA, which then forwarded them to the ASIC. With the loopback bitstream having been loaded into the eFPGA, the PRBS frames were sent back (also known as "loopback") to the KCU105's FPGA, which, in turn, forwarded them back to the software. Upon receiving the frames, the software verified the absence of bit errors in the frames.

This test was conducted using the same clock frequencies steps from Section~\ref{sec:cmos28_power_draw}, ranging from 10 MHz to 250 MHz. At each clock frequency step, the test ran for approximately 10 minutes, during which the software monitored for bit errors in the received frames compared to the frames it had sent. Throughout these 10-minute intervals, frames were continuously sent by the software, with the software transmission rate being constrained by the ASIC's PGP link rate, corresponding to the ASIC's clock frequency (e.g., a 250MHz clock resulted in a 250 Mbps link rate). Throughout all the steps of clock frequency tested, no bit errors were detected.


%% file: ml_on_eFPGA.tex
\section{Application for Machine Learning-based At-Source Processing}
\label{sec:ml_on_eFPGA}


In light of the successful eFPGA design and fabrication, scientific applications of the technology can be considered.
A variety of HEP readout tasks can benefit from reconfigurable logic within an ASIC.
For example, silicon pixel detectors detect the passage of charged particles and are commonly the first layer of high-energy collider detectors after the beam pipe, making them subject to the highest particle fluxes and radiation doses of any subsystem.  
The new all-silicon ATLAS Inner Tracker (ITk), to be installed in preparation for the HL-LHC, comprises 1.4 billion pixels with sizes down to 250 $\mu$m$^2$, resulting in an overall data rate of $\mathcal{O}$(10) Tb/s~\cite{itk_tdr}. 
However, the large majority of this data rate goes to the recording of "pileup" particles, which arise not from the hard-scattering proton-proton collisions, but from soft adjacent collisions that are less likely to contain physics processes of interest. 

Future detector designs for an $e^+e^-$ Higgs factory could benefit from reduction of the data rate going off the detector to minimize material budget associated to data cabling or mitigate the computational power required for trigger decisions. 
Further benefits of at-source ML-based processing will come to bear for detectors at future 10 TeV parton-center-of-mass ``discovery" machines, such as the Future Circular Collider (hh), where data rates are expected to near 1 exabyte per second with unparalleled radiation doses~\cite{Benedikt:2651300}.

\begin{figure}[!htbp]
\centering
   \includegraphics[width=0.75\textwidth]{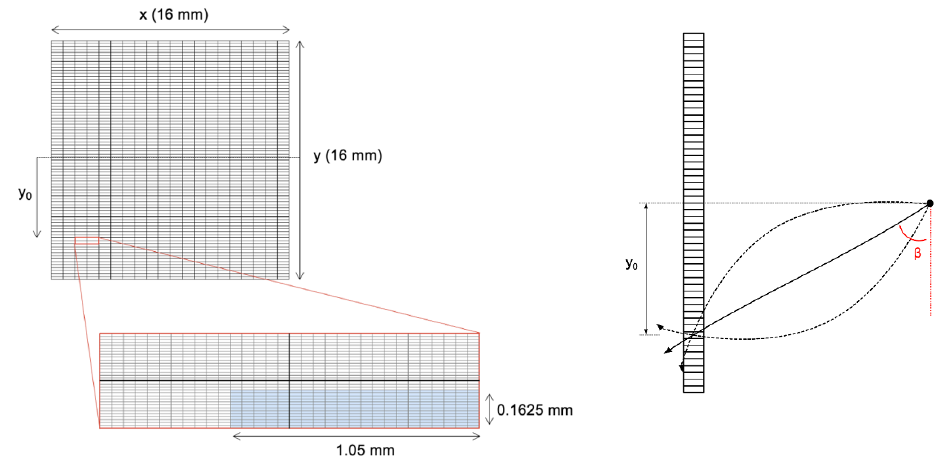}
    \includegraphics[width=0.6\textwidth]{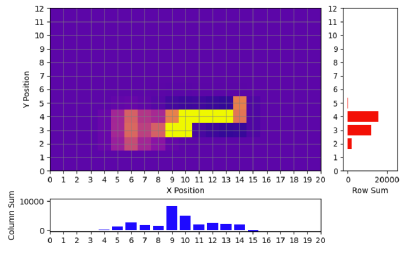}
   \caption{A diagram of the smart pixel sensor, where the blue shaded part corresponds to the local area within the sensor over which a single particle ionizes the silicon (top — Reproduced from~\cite{smartpixels}. \copyright~The Author(s).
Published by IOP Publishing Ltd. CC BY 4.0.), and an example smart pixel track represented as deposited charge across the 2D pixel grid (bottom). \label{fig:smartpixel}}
\end{figure}

The ``smart pixel'' collaboration provides a simulated dataset of particles propagated through a futuristic pixel sensor, which can be utilized to study ML-based readout~\cite{smartpixels_dataset}.
This dataset comprises about 550,000 fitted tracks originating from high-energy pions collected by the Compact Muon Solenoid (CMS) experiment and propagated through a futuristic pixel detector. 
This detector features sensors composed of a 21x13 pixel array with a 50 x 12.5 $\mu$m pitch, located at a radius of 30 mm from the beamline within a magnetic field of 3.8 T. 
Each track is represented as a sequence of eight deposited charge arrays in the (x, y) pixel dimensions, at time intervals of 200 ps.
Figure~\ref{fig:smartpixel} illustrates a diagram of a pixel sensor from this dataset, and an example signal from a charged particle track passing through it. 

Information about deposited charge in the pixel over time can be used to distinguish high-momentum particles from the hard-scatter proton interaction from low-momentum particles arising from pileup. 
High-momentum particles are less curved in the magnetic field, therefore traversing fewer pixels compared to pileup particles.
It is important to note that the x-profile (sum over pixel columns) runs parallel to the magnetic field and is thus not sensitive to particle momentum. In contrast, the y-profile (sum over pixel rows) is sensitive to the track's incident angle, and thereby its momentum.

Previous work has shown the feasibility of using ML to classify high-momentum tracks (or signal) from pileup (or background) based on the pattern of charge distribution across the sensor over time, enabling real-time data rate reduction by rejecting pileup tracks at the sensor level.
This pileup classification model has also been demonstrated to be feasibly integrated into an ASIC design~\cite{smartpixels}.
Another approach for on-chip ML involves performing \textit{regression} on low-level data to determine particle energy and angular information at the front-end, thus reducing the volume of information that needs to be transmitted~\cite{smartpixels_regression}.
Both approaches can significantly reduce the amount of data required to be transferred off-detector, leading to reductions in cabling and computational power requirements in the trigger system.
However, these methods currently depend on dedicated ASIC designs, which cannot be reconfigured to any other algorithmic operation, suggesting a potential for further advancement with an eFPGA-based approach.

The 28nm eFPGA was used as a proof-of-concept for reconfigurable logic in pixel readout, via the application of small ML-based methods for pileup classification using the smart pixel dataset. 
The specific task was to output a probability that a particle passing through the detector/tracker has a transverse momentum \(p_T > 2 \, \text{GeV}\), indicating it is likely to be pileup and thus should not be retained for further offline analysis.

An initial attempt was to design a simple Neural Network (NN) with two or three fully connected layers. Despite utilizing a few nodes per layer, this shallow NN required over 6,000 LUTs, significantly exceeding the capacity of the 28nm eFPGA ASIC. 
As an alternative, a Boosted Decision Tree (BDT) model was considered, known for its fast training, robust performance, and resource efficiency. Since a BDT primarily relies on comparison operations and thresholds, which can be directly embedded into the LUT logic and flip-flop registers, it does not require any block RAM or DSP resources, making it better suited to the resource-constrained hardware scenario. 

\begin{figure}[!htbp]
\centering
   \includegraphics[width=1.0\textwidth]{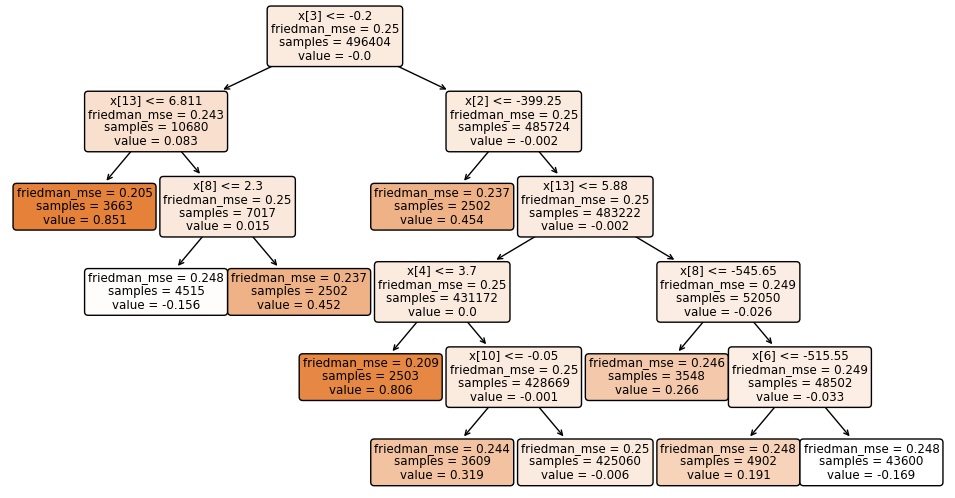}
   \caption{A diagram of the single tree BDT model used for proof-of-concept synthesis to the 28 nm eFPGA. \label{fig:BDT_model}}
\end{figure}

The BDT models each track as a 1D array of 14 values, 13 of which are the \(y\)-profile for each pixel summed over time, and the distance of the pixel from the interaction point \(y_0\), and outputs a probability that the track has \(p_T < 2\) GeV.
The full smart pixel dataset is split into orthogonal train and test sets with a ratio of 80\%:20\%. 
As described in Section~\ref{sec:cmos28nm}, the current 28nm eFPGA has a very small logical capacity, with only 448 LUTs. 
Given these stringent resource constraints, this model consists of a single tree with a depth of 5 and uses gradient boosting with the \textsc{scikit-learn} package~\cite{sklearn}, as shown in Figure~\ref{fig:BDT_model}. 

The performance of the BDT for the pileup classification task is evaluated using the test dataset.
The background rejection, defined as the fraction of pileup tracks (with truth $p_T < $ 2 GeV) that are correctly classified as background and thus rejected, is evaluated for a given signal efficiency, defined as the fraction of tracks with truth $p_T > 2$ GeV that are correctly classified as signal. 
The performance of the classifier is determined by setting a threshold for the BDT output above or below which a track is classified as signal or background, respectively.
Before quantization, a background rejection of 4.35\% is achieved for a signal efficiency of 97.53\% for a classification threshold of $0.4922$. 
After synthesis with quantization using \texttt{ap\_fixed}<28,19>, the performance of the model under varying classification thresholds is shown in Table~\ref{tab:performance_quan}.
Quantization can lead to slight discrepancies in performance for a fixed threshold. 
\begin{table}[!htbp]
    \centering
    \begin{tabular}{c | c | c}
        Signal Efficiency & Background Rejection & Classification threshold\\
        \hline\hline
        96.4\% & 5.8\% & 0.4953\\
        97.8\% & 3.9\% & 0.4922
    \end{tabular}
    \caption{Performance of the synthesized and quantized BDT model under different classification thresholds, defined by the BDT output.} \label{tab:performance_quan}
\end{table}

While the performance of this model is limited by the small logical capacity of the hardware, its ability to successfully perform the classification task and fit the resource limitations enabled a full configuration test on the 28nm eFPGA.
High-level synthesis of the software algorithm was performed using \textsc{Conifer}~\cite{hls4ml_bdt}, an open-source framework for generating firmware configuration files from software-based BDT algorithms. 
A full hardware synthesis of the pileup classification BDT was achieved, including the quantization of split threshold values in BDT and pruning to accommodate the BDT within stringent resource constraints.
In addition, the hardware output was successfully compared to the expected output from the quantized software version of the model.
This process involved the synthesis from C to Verilog firmware and the simulation of the hardware response. 
A single tree was synthesized into a single decision function module in RTL, featuring only 9 threshold parameters and 14 inputs. 
The entire module utilized 294 LUTs, fitting within the confines of the current 28nm eFPGA design. 
The compiled bitstream was then loaded into the eFPGA, and a test of the BDT evaluation was performed using the full smart pixel dataset of approximately 550,000 tracks. 
The model output when run in hardware was found to be 100\% accurate with respect to the expected output from the quantized software version of the synthesized model, with an operational runtime of less than 25 ns in simulation. 

These results provide a proof-of-concept for the use of ML on eFPGAs, confirming the ability to use the open-souce FABulous framework for eFPGA designs and opening the door to a variety of future applications in the sciences. 
However, further development is needed to enable eFPGA technology for readout in future collider detectors.
The small logical capacities of the ASICs described here are not feasible for data processing at future HEP particle detectors, which need to read out $\mathcal{O}$(10$^5$) pixels with very good signal efficiency and better background rejection than what is provided by the BDT described here. 
A next-generation eFPGA with a larger logical capacity could enable the study of higher-performance models and a variety of algorithms to fully exploit the configurability of the ASIC. 
Further studies into power consumption will also be needed to determine whether the eFPGA can be sufficiently efficient for the spatially constrained detector environment. 
Additionally, any readout ASIC in a collider experiment will need to be insensitive to radiation-induced issues such as single-event effects. 
The implementation of triple modular redundancy (TMR) in FABulous could open up a broad usage of eFPGAs in collider experiment readout scenarios. 

%% file: summary.tex
\section{Summary}
\label{sec:Summary}

This work describes the development of eFPGA technology using the FABulous open-source design framework. 
Two eFPGAs were designed and fabricated using 130 nm and 28 nm CMOS technologies respectively, and were tested for programmability, power consumption, and performance. 
The reconfigurability of the eFPGA makes it an excellent candidate for readout of silicon pixel detectors in high-energy collider experiments, which can benefit from ML-based filtering or feature extraction to reduce data rates. 
A proof-of-concept BDT is designed to classify and reject pile-up tracks ($p_T < 2$ GeV) using a simulated dataset of high-energy pions passing through a generic pixel sensor. 
Configuring this BDT model on the eFPGA achieved 100\% accuracy compared to the golden results. 
Future work will focus on the development of the FABulous framework, co-design of high-performance sensor processing algorithms, a larger eFPGA design with higher logic density for more complex processing, and dedicated radiation tolerance measures in order to accommodate realistic future collider experiments readout scenarios.